\documentclass[fleqn,twoside,twocolumn,nofootinbib]{revtex4} 
\usepackage{ujp} 
\begin{document}
\title[NOVEL PHOTOLUMINESCENCE-ENHANCING SUBSTRATES]
{NOVEL PHOTOLUMINESCENCE-ENHANCING SUBSTRATES FOR IMAGE FORMATION\\
OF
BIOLOGICAL OBJECTS}%
\author{G.I.~Dovbeshko}
\affiliation{Institute of Physics, Nat. Acad. of Sci. of Ukraine}
\address{46, Prosp. Nauky, Kyiv 03028, Ukraine}
\author{O.M.~Fesenko}
\affiliation{Institute of Physics, Nat. Acad. of Sci. of Ukraine}
\address{46, Prosp. Nauky, Kyiv 03028, Ukraine}
\author{V.V.~Boyko}%
\affiliation{Institute of Physics, Nat. Acad. of Sci. of Ukraine}
\address{46, Prosp. Nauky, Kyiv 03028, Ukraine}
\author{V.F.~Gorchev}
\affiliation{O.V.~Palladin Institute of Biochemistry, Nat. Acad. of Sci. of Ukraine}%
\address{9, Leontovych Str., Kyiv 01601, Ukraine}%
\author{S.O.~Karakhin}%
\affiliation{O.V.~Palladin Institute of Biochemistry, Nat. Acad. of Sci. of Ukraine}%
\address{9, Leontovych Str., Kyiv 01601, Ukraine}%
\author{N.Ya.~Gridina}
\affiliation{A.P.~Romodanov Institute of Neurosurgery, Academy of
Medical
Science of Ukraine}%
\address{32, Manuil's'kyi Str., Kyiv 04050, Ukraine}%
\author{V.S.~Gorelik}%
\affiliation{P.N.~Lebedev Physical Institute, Russian Academy of Sciences}%
\address{53, Lenin Ave., Moscow 117924, Russia}%
\author{V.N.~Moiseenko}
\affiliation{Dnipropetrovsk National University}%
\address{72, Gagarin Ave., Dnipropetrovsk 49050, Ukraine}%

\udk{???} \pacs{87.15.mq, 78.47.jd,\\[-3pt] 78.55.-m, 42.70.Qs, 87.64.mk}

\razd{\secix}
\setcounter{page}{732}%
\maketitle



\begin{abstract}
The use of photonic crystals, which were fabricated on the basis of
synthetic opals, as substrates for the luminescence microscopy
of biological objects has been shown. The spatial distributions of
the photoluminescence by DNA clusters excited by 365-nm ultra-violet
irradiation on opal surfaces and rough gold substrates have been
studied. With the use of blood cells as an example, a possibility for the visualization
of biological objects in the case where the nanostructure elements of
synthetic opals are applied as labels and image amplifiers has been
demonstrated.
\end{abstract}

\section{Introduction}

It is known that, in order that some biological problems be
solved, the contrast images of cells and their components have to
be obtained. The difficulty consists in that the cells, which are
transparent objects, are badly visible with an optical microscope.
Various approaches are used to tackle this problem, and the usage
of various dyes, quantum dots, and metallic nanoparticles as
luminescent labels \cite{1} can be classed as such. Being bound to
various molecular structures, those labels luminesce under the
action of laser radiation and visualize the site, where the cell
is connected with the label. In particular, 4$^{\prime
}$,6-diamidino-2-phenylindole (DAPI) either \textquotedblleft
dyes\textquotedblright\ cell nuclei or stimulates complexes with
DNA to luminesce in the blue spectral range; the green fluorescent
protein (GFP) is used as a protein label, including the
cytoskeleton; and so on \cite{2}. Autoluminescence of cell
components is very weak and practically is not observable even if
a confocal microscope is applied.

An alternative approach includes the creation of special optical setups for
visualizing the transparent objects, with the confocal and phase-contrast
microscopies being an example \cite{3}. The combination of those two
approaches allows good results to be obtained as well.

To visualize cells and their components, we proposed to use special
substrates, which play the role of optical elements and, simultaneously,
luminescent labels. Confocal microscopy is known to enable the detection of
luminescent microobjects with the help of an optical microscope, the
aperture of which is located before a detector and ensures the
registration of photoluminescence (PL) only from the objects located
immediately in the focal plane. In such a manner, the three-dimensional images
of analyzed objects can be created. The method of confocal microscopy
provides a high contrast, a large depth of field, and a capability to scan
specimens layer-by-layer. This circumstance allows three-dimensional images
of individual cells, tissue sections, and small organisms to be created,
which has predetermined a wide dissemination of the method in medical and
biologic applications.

Earlier \cite{4}, synthetic opals were suggested to be used as
matrices filled with the analyzed substance to enhance the PL
intensity of DNA. To excite PL in DNA (its quantum yield at room
temperature is less than $10^{-4}$), various dyes are used, such as
ethidium bromide, acridine orange, and others \cite{5,6}, whereas
the application of styryl-cyanine dyes to the DNA allows one to
obtain the two-photon luminescence in the visible spectral range
\cite{7}. Earlier \cite{4}, we suggested synthetic opal to be used
to enhance the luminescence of DNA near 350~nm. We also demonstrated
there a possibility to obtain the luminescence from DNA molecules
located on the surface of synthetic opal (without a temperature
decrease and a dye application) and to visualize DNA clusters by
exciting them at a wavelength of 365~nm. In this work, we reported
our experimental data concerning the usage of photonic crystals
(PCs) created on the basis of synthetic opals and rough gold
substrates for the imaging of biological objects such as DNA
clusters and blood cells.

\section{Materials and Methods}

The application of a confocal setup gives rise to the image
contrast enhancement, because a small-size apperture is applied so that the
\textquotedblleft extraneous\textquotedblright\ light emitted by the
neighbor points of the analyzed object does not arrive at a detector. The
\textquotedblleft cost\textquotedblright\ of a contrast enhancement is the
necessity to use rather complicated schemes for the scanning of a specimen
or light beam, which increases the time required to obtain complete
information about the examined object.

According to the Rayleigh criterion for the resolution limit (an
intensity reduction by 26\% from its maximum value) in the case of a
confocal microscope, $r_{\mathrm{conf}}=0.44\frac{\lambda }{n\sin
\theta }=0.88\frac{{\lambda }^{\prime }}{D}F$, where ${\lambda
}^{\prime }=\lambda /n$. At the same time, for a conventional
optical microscope, $r_{\mathrm{resel}}=0.61\frac{\lambda }{n\sin
\theta }=1.22\frac{{\lambda }^{\prime }}{D}F$. Hence, the resolution
of a confocal microscope is only 1.4 times better than that of an
optical one \cite{8}. Therefore, the main advantage of a confocal
microscope is not an increase of the resolution, but a substantial
increase of the contrast when the image is formed.

Our experiments were carried out on a confocal laser scanning microscope
Carl Zeiss LSM-510 META equipped with an objective Plan-Neofluar 40x/0.6
Korr. To obtain object images with this microscope, we used lasers with
wavelengths of 405, 458, 488, and 633\textrm{~nm}. Object images were
registered with the help of a digital camera AxioCam. PL excitation was
carried out with the use of an ultra-violet lamp HBO~100 and applying blue
(FSet01~wf), green (Fset10~wf), and red (Fset20~wf) filters. Images in
the visible range and in the \textquotedblleft
transmission\textquotedblright\ geometry were obtained using a halogen lamp.
The scanning rate depends on the resolution, so it was equal to 1/5 sec per
layer for a microscope LCM-510 at a resolution of $512\times 512$. The
maximum resolution was $2048\times 2048$ \cite{9}.

To estimate the PL spectral density, we scanned a specimen over its surface
in the \textquotedblleft Lambda scan\textquotedblright\ mode, which allowed
the PL spectrum to be registered with the use of a line of tiny photodetectors
with a spectral resolution of 10.7\textrm{~nm}. For the quantitative analysis,
we used the ROI (Region of Interest) feature, which enabled us to obtain the
plot of the PL intensity vs the wavelength, with the
dependence being averaged over the selected region.

\section{Fabrication of Specimens}

Specimens of synthetic opal were fabricated within the method of natural
globule sedimentation, by evaporating the reaction mixture at a given
rate. For the synthesis of silicon dioxide globules, the modified Stober
method \cite{10} was applied. The molar ratio between the components in the
reaction mixture was
$\mathrm{NH}_{4}\mathrm{OH}$:$\mathrm{H}_{2}\mathrm{O}$:$\mathrm{C}_{2}\mathrm{H}_{5}\mathrm{OH}$:$\mathrm{Si}(\mathrm{OC}_{2}\mathrm{H}%
_{5})_{4}=0.76$:$18$:$11$:$0.14$. After the sedimentation, the
obtained crystals were annealed at a temperature of 800$~^{\circ
}\mathrm{C}$ to remove chemically bound water and the remnants of
organic compounds.

In our researches, we also used rough gold substrates. A thin film of gold
was obtained by the thermal deposition of gold (99.999\%) in vacuum onto a glass
substrate (TF-1, $20\times20$~mm) covered with a Cr layer. Provided that the
Cr layer was 10--20$~\mathrm{\mathring{A}}$ in thickness, the thickness of
gold film was 300--350~\AA\, and its roughness was about 50~\AA\ \cite{11}.

A suspension of blood cells in a 0.9\%-solution of sodium chloride was
prepared. Venous blood was taken from an elbow vein, and heparin was
added (0.1~ml per 10~ml of blood). Blood cells were centrifuged at a rate of
1500~rpm for 10~min. Then blood plasma was removed, and the cell
precipitate was diluted with the equal volume of an isotonic sodium chloride
solution. The test tube with a specimen was stirred up and centrifuged
once more. Then, the precipitate was removed from the test tube, by preserving
0.5~ml over the volume occupied by the cells. At last, the cells
were deposited onto a glass substrate or the surface of synthetic opals for
their study with the confocal microscope.

\section{Results and Their Discussion}

Since the surface of a PC created on the opal basis has a domain
structure and is iridescent, the first stage of our research
consisted in optical studies of the initial opal surface by using
the confocal microscopy. In Fig.~1, the regions with ordered
structure (domains) from 40 to 250~$\mu \mathrm{m}$ in size observed
on the opal surface are exhibited. The luminescence of the PC
surface was detected, when it was excited by UV light with a
wavelength of 365~nm from a mercury lamp. The registered glow
produced by inhomogeneities on the opal surface allowed us to
distinguish separate domains well.

\begin{figure}
\includegraphics[width=\column]{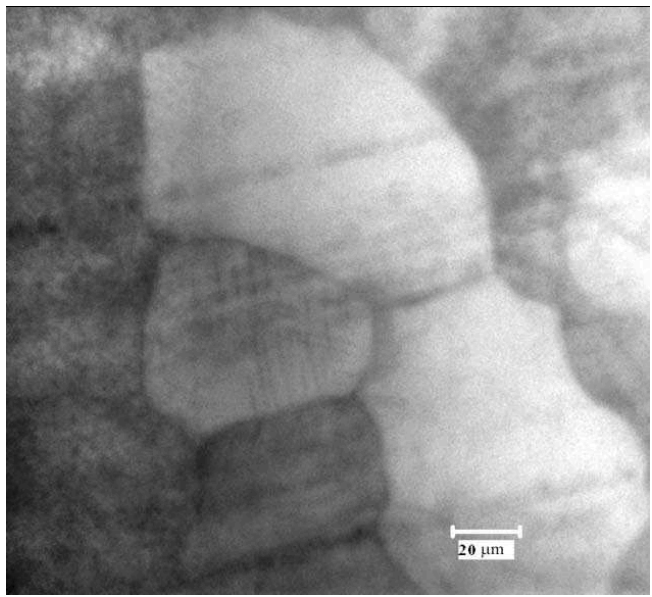}
\vskip-3mm\caption{View of the synthetic opal surface under its
excitation with an ultra-violet lamp. The fluorescence is registered
with the use of a blue optical filter of the type FSet01  }
\end{figure}

The researches showed \cite{4,12,13} that the initial PC
demonstrates PL with two intensive maxima at about 400 and
500~nm, and several weak bands in the interval of 600--800~nm. In
work \cite{14}, PL produced by a synthetic opal was associated with
the presence of defects and impurities. The band with a maximum
at a wavelength of 523~nm is related to the surface states of
$\equiv$Si--H (a transition energy of 2.37~eV), whereas the band
with maxima at 652 and 692~nm to the bulk and surface states of
$\equiv$Si--O (transition energies of 1.9 and 1.79~eV,
respectively). The nature of the band at about 400~nm was discussed in
\cite{12}. This band is associated with the
presence of various impurities such as zinc, calcium, sodium, iron, and zirconium
oxides, which get into specimens in the course of crystal
fabrication. The fraction of these impurities is less
than $10^{-5}.$ The PL intensity for the initial synthetic opal changed
from point to point under the scanning over its surface (see Fig.~2).
For the specimens with a well-ordered structure, the variation of PL
intensity over the surface was observed with an identical
coefficient for various frequencies. For the specimens with a less
ordered structure, the luminescence intensity changed differently at
different frequencies. Such a dependence of the synthetic opal
luminescence intensity on the surface site can be connected with the
inhomogeneity of a specimen, the presence of a domain structure, the
variation in the spectral position of the energy gap, and the
non-uniform distribution of impurities. Since the size of domains in
PCs is comparable with the wavelength of light in the visible range,
one may expect a local increase of the electromagnetic field for the
impurity-induced PL, so that a signal even from a very small number
of impurities could be registered \cite{15,16}. The energy gap of PC
reveals itself in a luminescence band; namely, a dip has to be
observed in the luminescence contour in the spectral region, where
the PC energy gap is located \cite{12}. For the studied crystals
with a globule size of 240~nm, the Bragg reflection should take
place at about 530~nm. In accordance with that, a \textquotedblleft
dip\textquotedblright\ was registered in the corresponding region of
the opal PL spectra (see Fig.~2).

\begin{figure*}
\includegraphics[width=15cm]{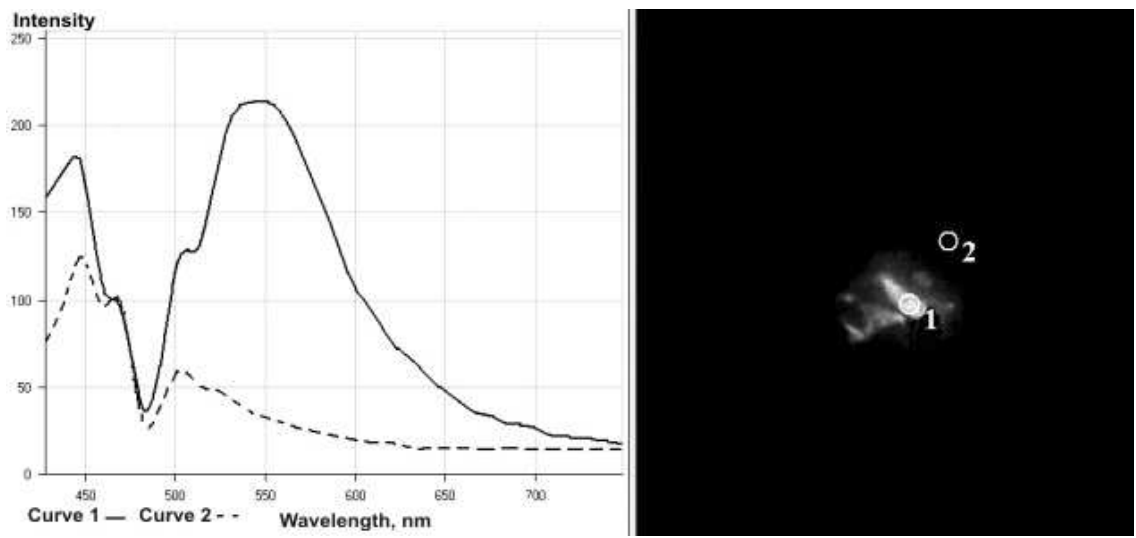}
\vskip-3mm\caption{(right panel)~Photoluminescence image of
a luminescing region on the surface of synthetic opal in a vicinity
of the defect obtained in the Lambda Scan mode. Photoluminescence was
excited with a laser at a wavelength of 405\textrm{~nm} and
registered in the range of 422--754\textrm{~nm} with the use of
Beamsplitter 405/488. The left panel exhibits the corresponding PL
spectra obtained from two regions on the opal surface marked on the
right panel  }
\end{figure*}

The main electron transition $S_{0}$--$S_{1}$ in a DNA molecule
corresponds to the electromagnetic radiation wavelength close to
260~nm. Having absorbed an ultra-violet quantum, the DNA molecule
transits from the ground state $S_{0}$ in that corresponding to
the first (resonance) singlet electronic term. This transition is
allowed by the selection rules for single-photon processes. Slightly
lower on the energy scale, there is located a triplet term, for
which the single-photon processes into the ground state are
spin-forbidden. Accordingly, a condition for the PL spectra in
DNA to be excited is an excess of the energy of an exciting radiation
quantum over the resonance electron transition energy; i.e. the
exciting radiation wavelength has to be shorter than or at least
comparable with 260~nm. After the excitation (owing to the single-photon
absorption) of a DNA molecule and its transition onto a singlet term,
conversion processes become possible, when the molecule transits
onto a triplet term characterized by a very small oscillator
strength \cite{17}. In addition, the conversion can take place from term
$S_{0}$ into a number of other terms with small oscillator strengths
connected with the electronic structure of nucleic bases composing
DNA. As a result, the main fraction of the electromagnetic energy
absorbed by a DNA molecule transforms into the energy of thermal
motion. In this connection, the quantum yield of PL for DNA
molecules is extremely low (less than $10^{-4}$), which makes the
observation of PL by a native DNA extremely difficult even if the
sources of exciting photons of short-wave electromagnetic radiation
with a wavelength shorter than 260~nm are used. In this connection,
the known results concerning the registration of PL spectra for the
native (label-free) DNA excited with short-wave radiation \cite{5,6}
were obtained only at low temperatures.

In order to enhance the intensity of PL by DNA molecules located on the
surface of a condensed medium, we proposed to use two new types of substrates:
a rough film of gold and a micro-structured surface of synthetic opal.

In Fig.~3, the image of DNA strands located on a rough gold surface is
depicted. The image was obtained by registering PL excited by ultra-violet
radiation (365~nm). The figure demonstrates that the objects are observed in
the form of strands 5--7~$\mu \mathrm{m}$ in length and about 1~$\mu
\mathrm{m}$ in thickness. A real thickness of DNA strands is known to equal tens
of nanometers. This conclusion was confirmed earlier by the results of
researches of the DNA microstructure with the use of atomic-force \cite{18} and
electron microscopies. An increase of the transverse dimensions of the discussed
object (to 1~$\mu \mathrm{m}$) at its observation by the PL method with the
help of a confocal microscope can be explained by a probable condensation of
DNA molecules with the assembling of microclusters on the gold surface. As
Fig.~3 demonstrates, the images of DNA strands are bright and clear enough.
The higher intensity of a PL signal from DNA molecules on the gold surface can
be associated with a manifestation of the known effect consisting in a huge (by
several orders of magnitude) amplification of the effective electromagnetic PL
field near gold nanoparticles. This effect is explained by the influence of
localized surface plasmons. On the other hand, the chemical processes of
electron transport from the metal to DNA molecules, which can give rise to
the formation of new electron states in the visible spectral range, are not
excluded as an alternative mechanism of PL intensity enhancement. In effect,
the latter mechanism corresponds to the formation of new fluorescent labels
in a vicinity of gold nanoparticles. As a result of the spontaneous label
formation in the \textquotedblleft gold--DNA\textquotedblright\ complex, PL
in DNA can be excited by radiation in the middle ultra-violet range
(365\textrm{~nm}).

\begin{figure}
\includegraphics[width=\column]{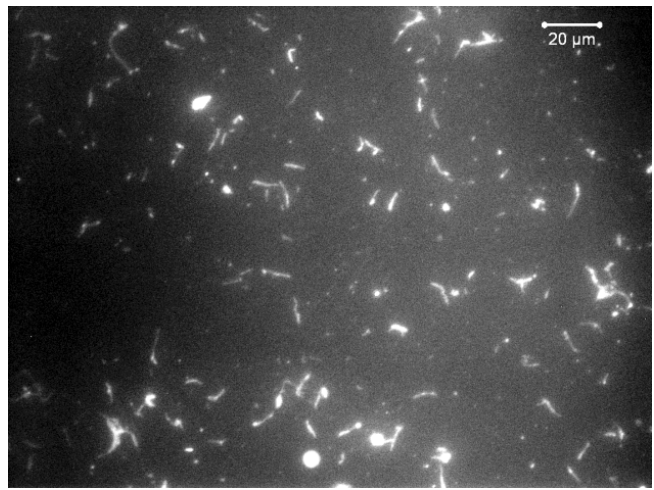}
\vskip-3mm\caption{Photoluminescence image of DNA molecules
deposited onto a rough gold substrate. The image was obtained with
the help of a confocal microscope and by exciting PL with the use of
ultra-violet radiation with a wavelength of 365\textrm{~nm}  }
\end{figure}

\begin{figure}
\includegraphics[width=6.7cm]{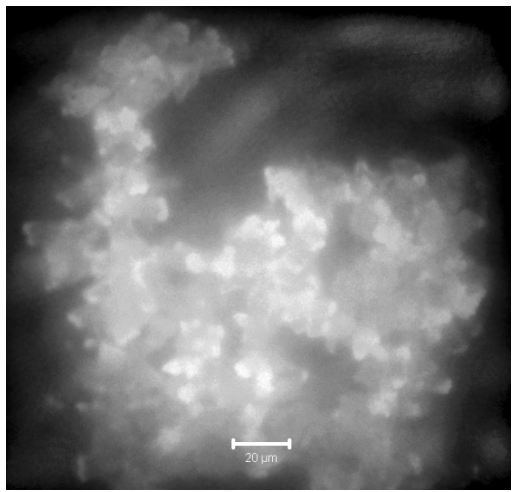}
\vskip-3mm\caption{Photoluminescence image of DNA molecules
deposited onto the surface of a photonic crystal. The image was
obtained with the help of a confocal microscope. PL
was excited by ultra-violet radiation with a wavelength of
365\textrm{~nm}  }
\end{figure}

In Fig.~4, the image of DNA microclusters deposited onto the PC surface is
shown. The image was obtained in the PL light generated at the excitation at
a wavelength of 365~nm. The figure demonstrates that the images of DNA
molecules are more smeared than those in the case where the rough gold
substrate was used. Nevertheless, a rather intensive PL signal is also observed
in this case, which allows us to obtain information on the spatial
distribution of DNA over the substrate surface. It is known that, if the
size of silica globules in a PC is 250~nm, the size of interglobular pores
is about 50~nm, and the diameter of channels that reach the PC surface is
close to 10~nm. Therefore, the smear of the images of DNA strands on the PC
surface can be explained by a \textquotedblleft sinking\textquotedblright\
of DNA molecules into the PC bulk through the interglobular channels.

The observed effect that the image intensity (in the PL light) of DNA
clusters located on the PC surface grows can be explained, similarly to the
case with rough gold substrates, by a local increase of the effective field
\cite{19}. However, now the effect takes place not as a result of the
plasmon effect but owing to the interference-induced amplification of
the effective field in the PC due to a growth of the density of photon states
in a vicinity of defects and the edges of the forbidden photon band \cite{15}.
The interference character of the field amplification can serve as an additional
reason for the smearing of the images of biological objects on the PC surface.

It should also be noted that nanostructured gold, as well as nanostructured
PCs, has a variety of electron transitions in the visible and near
ultra-violet ranges of the spectrum \cite{20,21}. They correspond to
energies of 2.4~eV (0.52~$\mu \mathrm{m}$), 2.5~eV (0.5~$\mu \mathrm{m}$),
1.9~eV (0.65~$\mu \mathrm{m}$), 2.2~eV (0.56~$\mu \mathrm{m}$), 3.2~eV
(0.3~$\mu \mathrm{m}$), 2.55~eV (0.49~$\mu \mathrm{m}$), and 3.0~eV (0.41~$\mu
\mathrm{m}$). The presence of these levels creates conditions for the
formation of corresponding labels in the complexes \textquotedblleft
DNA--substrate defect\textquotedblright\ and for the emergence of a PL signal when
the latter are excited by radiation with a wavelength of 365~nm.

\begin{figure}
\includegraphics[width=\column]{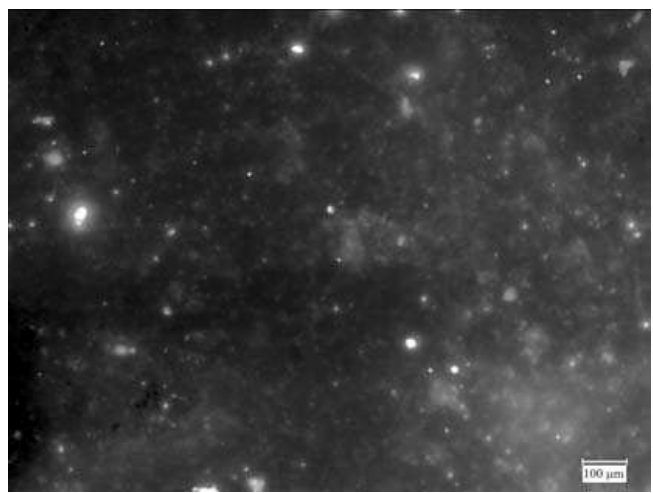}
\vskip-3mm\caption{Photoluminescence image of blood plasma deposited
onto the surface of a photonic crystal. The image was obtained with
the help of a confocal microscope, by exciting PL with the use of
ultra-violet radiation with a wavelength of 365\textrm{~nm}
and applying a blue optical filter FSet01. \textquotedblleft
Red\textquotedblright\ and \textquotedblleft blue\textquotedblright\
luminescent points correspond to different biological objects  }
\end{figure}

In addition,
if the PC substrate is microstructured, the emergence of new chemical bonds
between DNA molecules and defects on the PC surface is possible analogously to
the case of inhomogeneities on the gold substrate. New
chemical bonds \textquotedblleft PC--DNA\textquotedblright\ play a role of
formed labels, which give rise to PL when being excited by radiation in the
middle UV-range (365~nm).

Figure~5 exhibits the images of blood plasma cells, and Fig.~6
presents the corresponding PL images of erythrocytes deposited onto
the surface of synthetic opal. As is seen from those figures, when
the microscope is adjusted to watch the surface layer of cells, the
glow is observed to be emitted directly from the cells rather than
the substrate surface. The images of biological objects presented in
Figs.~5 and 6 enable the dimensions of cells and their shapes to be
estimated. While analyzing the PL spectra from definite object
regions, it is possible to obtain quantitative information
concerning the parameters of secondary radiation from cell
components. One can distinctly observe a concave at the center of an
erythrocyte cell. It should be noted that, when depositing blood
cells onto the cover glass, none of their PL images were obtained.
At the same time, when a PC was used as a substrate, the observable
(in the PL light) images differ from one another not only by their
shapes, but also by their spectra.

\section{Conclusions}

Hence, the following results were obtained in this work.

1.~When DNA molecules are deposited onto a rough gold substrate and the
exciting ultraviolet radiation with a wavelength of 365~nm is used, it is
possible to form a distinct image of
photoluminescent DNA clusters observed with a confocal microscope as strands several micrometers in
length. The observed width of strands substantially exceeds the real
transverse size of DNA molecules. This result is explained by the resolution
limit of a confocal microscope (the luminescence mode).

2.~When DNA molecules are deposited onto a nanostructured surface of
a photonic crystal, the PL image of DNA clusters can also be formed with the
help of a confocal microscope, by applying exciting ultraviolet radiation with a
wavelength of 365~nm. However, the images of DNA molecules are
smeared and less intense in this case. The smearing of the photoluminescence
image of DNA can be explained by the \textquotedblleft
sinking\textquotedblright\ of DNA molecules into the PC bulk through the
channels, whose diameter is close to 10\textrm{~nm}.

3.~The specific feature of the method proposed is the fact that the application of
a photonic crystal makes it possible to register the photoluminescence images
of biological objects with large dimensions (more than 1~$\mu \mathrm{m}$),
which do not \textquotedblleft sink\textquotedblright\ through the PC pores.
It is also true for clusters composed of a large number of molecules to be
studied, which fill the whole PC volume \cite{4}. Therefore, we can
conclude that the application of PCs as substrates is expedient for the
formation of the photoluminescence images of biological objects characterized by
a rather big size (cells, viruses, {\it etc.}). In this case, the usage of
substrates made of PC turns out considerably more efficient than the usage
of a cover glass.

For the further usage of the method, in which PCs created on the basis of
synthetic opals are applied as substrates, it seems expedient to fabricate
opals with smaller dimensions of globules, which would provide the
corresponding reduction of channel dimensions to about 1~nm. Moreover, it is
reasonable to excite PL in native DNAs with the help of
short-wave radiation. In particular, semiconductor light-emitting diodes
with a wavelength of 280\textrm{~nm} or the fourth harmonic of a YAG:Nd$^{3+}$
laser with a generation wavelength of 266\textrm{~nm} can be used for this
purpose.

\begin{figure}
\includegraphics[width=\column]{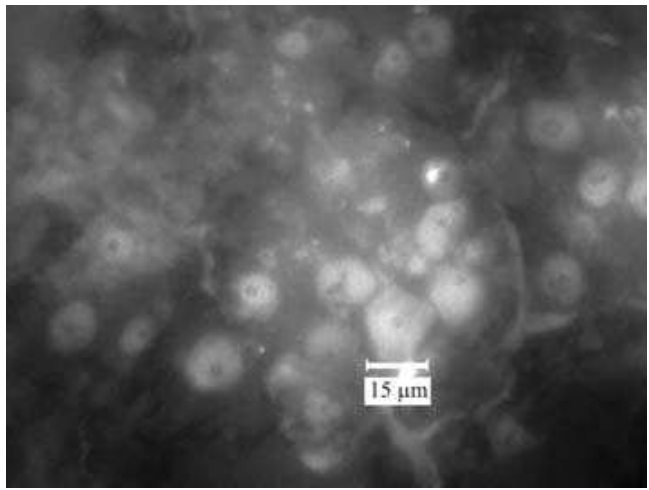}
\vskip-3mm\caption{Photoluminescence image of erythrocytes deposited
onto the surface of a photonic crystal. The image was obtained with
the help of a confocal microscope, by exciting PL with the use of
ultra-violet radiation with a wavelength of 365\textrm{~nm} and
applying a blue optical filter FSet01}
\end{figure}

\vskip3mm
 The work was sponsored by the Ukrainian-Russian project
(2012--2013), the project of the Science and Technology Center in
Ukraine N~5525 (2012--2013), and the Ukrainian-German project N~M366
(2011--2012). The authors are also grateful to the Ministry of
Education and Science of the Russian Federation (state contract
N~16.513.11.3116) and the Russian Foundation for Basic Research
(projects N~10-02-00293, 11-02-00164, and 11-02-12092).

\rezume{%
 НОВІ ПІДКЛАДКИ, ЩО ПОСИЛЮЮТЬ СИГНАЛИ\\ ФОТОЛЮМІНЕСЦЕНЦІЇ, ДЛЯ
 ФОРМУВАННЯ\\ ЗОБРАЖЕНЬ БІОЛОГІЧНИХ ОБ'ЄКТІВ} {Г.І. Довбешко, Є.М.
 Фесенко $ ^ {1} $, В.В. Бойко, В.Ф. Горчев,\\ С.О. Карахін, Н.Я. Гридіна,
 В.С. Горелик, В.Н. Моісеєнко} {В роботі показано можливість
 використання фотонних кристалів, створених на основі штучних
 опалів, як підкладок для люмінесцентної мікроскопії
 біологічних об'єктів. Вивчено просторовий розподіл
 фотолюмінесценції кластерів ДНК на поверхні опалів і на
 шорсткуватих золотих підкладках при дії ультрафіолетового
 випромінювання з довжиною хвилі 365 нм. На прикладі зразків клітин крові
 продемонстровано можливість візуалізації біологічних об'єктів
 при використанні елементів наноструктури штучних опалів у
 ролі міток і підсилювачів зображення.}

\end{document}